\documentclass[11pt]{article}
\usepackage{amsfonts,amsthm,amsmath,amssymb}

\newcommand{\Comment}[1]{{}}

\def\a{\alpha}
\def\b{\beta}

\def\d{\partial}

\newcommand{\be}{\begin{equation}}
\newcommand{\bea}{\begin{eqnarray}}

\newcommand{\ee}{\end{equation}}
\newcommand{\eea}{\end{eqnarray}}

\begin{document}
\rightline{TIT/HEP-596}
   \vspace{2.8truecm}

\vspace{15pt}


\centerline{\LARGE \bf On IR solutions in Horava gravity theories }  \vspace{2truecm}
\thispagestyle{empty} \centerline{
    {\large \bf Horatiu Nastase}\footnote{E-mail address: \Comment{\href{mailto:nastase.h.aa@m.titech.ac.jp}}{\tt nastase.h.aa@m.titech.ac.jp}}}

\vspace{1cm}
\centerline{{\it  Global Edge Institute, Tokyo Institute of Technology,}}
\centerline{{\it Meguro, Tokyo 152-8550, Japan}}

\vspace{3truecm}

\thispagestyle{empty}

\centerline{\bf ABSTRACT}

\vspace{1truecm}

\noindent
In this note we search for large distance solutions of Horava gravity. In the case of the "detailed balance"
action, gravity solutions asymptote to IR only above the cosmological constant ($\sim$horizon) scale. However, if one 
adds  IR dominant terms $\alpha R^{(3)}+\beta \Lambda_W$, one can recover general relativity solutions on 
usual scales in the real Universe, 
provided one fine-tunes the cosmological constant, reobtaining the usual cosmological constant 
problem. We comment on pp wave solutions, in order to gain insight into the relativistic properties of the 
theory.

\vspace{.5cm}

\setcounter{page}{0}

\newpage

\Comment{
\begin{flushright}
TIT/HEP-XXX
\end{flushright}
\vskip.5in

\begin{center}

{\LARGE\bf On IR solutions in Horava gravity theories }
\vskip 1in
\centerline{\Large Horatiu Nastase}
\vskip .5in

\end{center}
\centerline{\large  Global Edge Institute}
\centerline{\large Tokyo Institute of Technology, Meguro, Tokyo 152-8550, Japan}

\vskip 1in

\begin{abstract}

{\large 
}

\end{abstract}

\newpage}

In a recent paper \cite{Horava:2009uw} (based on the earlier work in \cite{Horava:2008ih}), 
Horava proposed a new theory, or rather a class of theories, for 
gravity in four dimensions, that is nonrelativistic in the UV, i.e. time plays a special role,
but should become relativistic and flow towards the 
usual general relativistic Einstein-Hilbert action in the IR. The important feature of this class of theories 
is its power counting renormalizability, at least around the flat space 
vacuum solution, so that one could maybe consider it 
a UV completion of general relativity. This does not preclude an embedding in string theory, just as the 
renormalizability of Yang-Mills theory does not preclude it being embedded in string theory, but it opens up new 
and interesting possibilities. A number of papers on Horava gravity have recently appeared 
\cite{Horava:2009if,Visser:2009fg,Takahashi:2009wc,Calcagni:2009ar,Kiritsis:2009sh,Kluson:2009sm,Lu:2009em,Mukohyama:2009gg,Brandenberger:2009yt,Nikolic:2009jg}.
Most of them examine issues of cosmology arising from it, and in \cite{Lu:2009em} solutions to Horava gravity were
analyzed. The purpose of this note is to see whether the simple "detailed balance" proposed by Horava
is enough, or rather one must add extra terms, and also to examine the relativistic properties of the action.

The ADM parametrization of a metric in 3+1 dimensions is
\be
ds^2=-N^2 dt^2 +g_{ij}(dx^i-N^idt)(dx^j-N^jdt)\label{admmetric}
\ee

Horava proposes that one takes a nonrelativistic gravity theory, where the functions $N,N_i$ and $g_{ij}$ are 
taken to be independent, thus such that there is no general covariance with respect to the 4d ADM metric 
(\ref{admmetric}).
In his theory, space and time scale differently in the UV, where one defines the theory, 
\be
\vec{x}\rightarrow b\vec{x};\;\;\;
t\rightarrow b^zt
\ee
with $z_{UV}=3$. The kinetic term of Horava's nonrelativistic gravity is 
\be
S_K=\frac{2}{\kappa^2}\int dt d^3x N (K_{ij}K^{ij}-\lambda K^2)
\ee
where, with the given scaling, both $\kappa$ and $\lambda$ are dimensionless couplings and where
$K_{ij}$ is the extrinsic curvature of the ADM foliation, 
\be
K_{ij}=\frac{1}{2N}(\dot{g}_{ij} -\nabla_i N_j-\nabla_j N_i)
\ee
For $\lambda=1$ we get the kinetic (time derivatives) part of the four dimensional Einstein-Hilbert
term
\be
S_{EH}=\frac{1}{16\pi G}\int d^4 x \sqrt{g}N(K_{ij}K^{ij}-K^2+R-2\Lambda),
\ee
where $R$ is the three dimensional Euclidean Ricci scalar for $g_{ij}$.

Horava then adds a 
nonrelativistic potential $S_V$. Purely for conditions of simplification, one can choose the potential 
satisfying the "detailed balance" condition, 
\be
S_V=\frac{\kappa^2}{8}\int dt d^3x \sqrt{g}N E^{ij}{\cal G}_{ijkl}E^{kl};\;\;\;\;
\sqrt{g}E^{ij}\equiv \frac{\delta W[g_{kl}]}{\delta g_{ij}}
\ee
where ${\cal G}_{ijkl}$ is the inverse of the De Witt metric, and $W$ is a 3d euclidean action. Horava then chooses 
\be
W=\frac{1}{w^2}\int \omega_3(\Gamma)+\mu \int d^3 x \sqrt{g}(R-2\Lambda_W),
\ee
where the first term is a Chern-Simons term, with the dimensionless coupling $w$, which defines Horava gravity in the
UV, and the second is a 3d 
Einstein-Hilbert term with a coupling $\mu$ of dimension 1 and a 3d cosmological constant $\Lambda_W$ of dimension
2, which is expected to be subleading in the UV 
(i.e., at short distances), and is just a simple choice of relevant deformation.

In the UV, this theory is power counting renormalizable, at least around the flat space (vacuum) solution. 
This action, after an analytical continuation, 
can be written as 
\bea
&& S_H=\int dt d^3 x ({\cal L}_0+{\cal L}_1)\cr
&& {\cal L}_0=\sqrt{g}N\Big[\frac{2}{\kappa^2}(K_{ij}K^{ij}-\lambda K^2)-\frac{\kappa^2\mu^2}{8(1-3\lambda)}
(\Lambda_WR-3\Lambda^2_W)\Big]\cr
&&{\cal L}_1=\sqrt{g}N\Big[-\frac{\kappa^2\mu^2}{32(1-3\lambda)}R^2+\frac{\kappa^2}{2w^4}Z_{ij}Z^{ij}\Big]\cr
&&Z_{ij}\equiv C_{ij}-\frac{\mu w^2}{2}R_{ij}
\eea
and $C^{ij}=\epsilon^{ijk}\nabla_k(R^j_l-\frac{1}{4}R\delta^j_l)$ is the Cotton tensor.

Its equations of motion were found in \cite{Lu:2009em,Kiritsis:2009sh} and are 
\bea
&&\frac{2}{\kappa^2}(K_{ij}K^{ij}-\lambda K^2)-\frac{\kappa^2\mu^2}{8(1-3\lambda)}
(\Lambda_WR-3\Lambda^2_W)-\frac{\kappa^2\mu^2}{32(1-3\lambda)}R^2+\frac{\kappa^2}{2w^4}Z_{ij}Z^{ij}=0\cr
&&\nabla_k[K^{kl}-\lambda K g^{kl}]=0\cr
&&\frac{2}{\kappa^2}E_{ij}^{(1)}-\frac{2\lambda}{\kappa^2}E_{ij}^{(2)}+\frac{\kappa^2\mu^2\Lambda_W}{8(1-3\lambda)}
E_{ij}^{(3)}+\frac{\kappa^2\mu^2(1-4\lambda)}{32(1-3\lambda)}E_{ij}^{(4)}-\frac{\mu\kappa^2}{8w^2}E_{ij}^{(5)}
-\frac{\kappa^2}{2w^4}E_{ij}^{(6)}=0\label{eom}
\eea
where the first two lines are the $N$ and $N_i$ equations, with the $N_i$ equations satisfied identically for static 
solutions, 
and the third is the $g_{ij}$ equation, where $E_{ij}^{(4,5,6)}$ come from ${\cal L}_1$ and $E_{ij}^{(1,2)}$
comes from the kinetic term in ${\cal L}_0$, while $E_{ij}^{(3)}$ comes from the 3d Einstein-Hilbert term.
Here 
\bea
E_{ij}^{(1)}&=& M_i \nabla_k K^k_j+N_j\nabla_k K^k_i -K^k_i\nabla_j N_k-K^k_j \nabla_i N_k-N^k\nabla_k K_{ij}\cr
&&-2NK_{ik}K_j^k-\frac{N}{2}K^{kl}K_{kl}g_{ij}+NK K_{ij}+\dot{K}_{ij}\cr
E_{ij}^{(2)}&=& \frac{N}{2}K^2g_{ij}+N_i \d_jK+N_j\d_i K -N^k (\d_k K)g_{ij}+\dot{K}g_{ij}\cr
E_{ij}^{(3)}&=&N\Big[R_{ij}-\frac{1}{2}Rg_{ij} +\frac{3}{2}\Lambda_W g_{ij}\Big]-\Big(\nabla_i \nabla_j-g_{ij}
\nabla_k \nabla^k\Big)N\cr
E_{ij}^{(4)}&=&NR\Big[2R_{ij}-\frac{1}{2}g_{ij}R\Big]-2\Big(\nabla_i\nabla_j-g_{ij}\nabla_k\nabla^k\Big)(NR)\cr
E_{ij}^{(5)}&=& 2\nabla_k \Big[\nabla_j(N{Z^k}_i)+\nabla_i(N{Z^k}_j)\Big]-2\nabla_k\nabla^k(NZ_{ij})-2\nabla_k
\nabla_l(NZ^{kl})g_{ij}\cr
E_{ij}^{(6)}&=&-\frac{1}{2} NZ_{kl}Z^{kl}g_{ij} +2NZ_{ik}{Z_j}^k-N(Z_{ik}{C_j}^k+Z_{jk}{C_i}^k)+NZ_{kl}C^{kl}g_{ij}\cr
&&-\frac{1}{2}\nabla_k\Big[N\epsilon^{mkl}(Z_{mi}R_{jl}+Z_{mj}R_{il}\Big]+\frac{1}{2}R^n_l\nabla_n\Big[N\epsilon^{mkl}
(Z_{mi}g_{kj}+Z_{mj}g_{ki})\Big]\cr
&&-\frac{1}{2}\nabla_n \Big[N{Z_m}^n\epsilon^{mkl}(g_{ki}R_{jl}+g_{kj}R_{il})\Big]\cr
&&+\frac{1}{2}\nabla_n\Big[\nabla_i \nabla_k(N{Z_m}^n\epsilon^{mkl})g_{jl}+\nabla_j\nabla_k(N{Z_m}^n\epsilon^{mkl})
g_{il}\Big]\cr
&& +\frac{1}{2}\nabla_l \Big[\nabla_i \nabla_k(NZ_{mj}\epsilon^{mkl})+\nabla_j\nabla_k(NZ_{mi}\epsilon^{mkl})\Big]
-\nabla_n\nabla_l\nabla_k(N{Z_m}^n\epsilon^{mkl})g_{ij}\cr
&&-\frac{1}{2}\nabla_n\nabla^n\nabla_k\Big[N\epsilon^{mkl}(Z_{mi}g_{jl}+Z_{mi}g_{il})\Big]\label{es}
\eea

In the IR, the terms of lowest dimension should dominate, so these are the terms in ${\cal L}_0$. Moreover, we 
see that for (quasi)static solutions, the three dimensional Ricci scalar $R$ and the cosmological constant $\Lambda_W$
will dominate. From the fact that the effective speed of light is $c=\kappa^2\mu \sqrt{\Lambda_W/(3\lambda -1)}/4$
and the cosmological constant is $\Lambda =3/2\Lambda_W$ we would hope that this would be enough to get the most 
general Einstein-Hilbert theory. But in fact, by 
comparing ${\cal L}_0$ and ${\cal L}_1$ we see that, at least naively, 
"the IR" where four dimensional 
general relativity is supposed to be recovered is where the three dimensional curvature of the solution 
$R$ is much smaller than the cosmological constant scale, i.e. $R\ll \Lambda_W\sim \Lambda$. However in that case, 
the cosmological constant $\Lambda$ will dominate over the four dimensional curvature $R^{(4)}$, so this would 
correspond in the present-day Universe
to the rather uninteresting case where we look at scales larger than the cosmological constant 
($\sim$horizon) size. We will study in the following whether this naive expectation is correct and whether it can be 
circumvented.

Fortunately however, the solution to this problem
is just to go beyond the overly simple "detailed balance" action to a more natural case. 
To the action $S_H$ satisfying detailed balance one can in principle add various other possible terms
that become subleading in the UV (relevant deformations), besides the ones proportional to $\mu$. In particular, 
the terms most dominant in the IR that one could add are
an arbitrary 3d Einstein term and a cosmological term, and they break detailed balance only softly
\cite{Horava:2009uw}. For ease of comparison with the above, we will 
write them as 
\be
\delta S_H=-\int dt d^3x \sqrt{g}N\frac{\kappa^2\mu^2}{8(3\lambda-1)}\Lambda_W
(\alpha R+\beta\Lambda_W)
\ee
so that in $S_H$ there is a term corresponding to $\alpha=1$ and $\beta=-3$. The addition of this term then 
only changes
\be
(\Lambda_WR-3\Lambda^2_W)\rightarrow ((1+\a)\Lambda_W R+(\beta-3)\Lambda_W^2)
\ee
in the first ($N$) equation of (\ref{eom}) and changes the expression for $E_{ij}^{(3)}$ in (\ref{es}) to 
\be
\tilde{E}_{ij}^{(3)}=N\Big[(1+\a)(R_{ij}-\frac{1}{2}Rg_{ij}) +\frac{3-\b}{2}\Lambda_W g_{ij}\Big]-
(1+\a)\Big(\nabla_i \nabla_j-g_{ij}\nabla_k \nabla^k\Big)N
\ee
and the rest remains unchanged.

In the UV region (at short distances), the theory will be nonrelativistic, and we must use an action like  
$S_H+\delta S_H$ with classical values for the couplings $\kappa,\lambda,w,\mu, \Lambda_W,\alpha,\beta$. 
However, if Horava gravity is to be of use for the real world,  
we want to ascertain what happens on usual scales. Then, we must consider scales that are much smaller than the 
cosmological constant scale (which is comparable with the horizon size), but much larger than quantum gravity scales, 
thus "in the IR". In that regime however, it is clear that the couplings will be heavily renormalized, away from 
their classical values. We expect that at least the 
renormalized coupling $\lambda_r$ should approach the Einstein-Hilbert 
value of $\lambda_{EH}=1$. Also, 
because of quantum corrections, we expect that even if we have $\alpha\simeq\beta\simeq 0$ 
in the UV, in the IR they should be nonzero.

Therefore in the rest of this note we will consider the theory "in the physical IR region", i.e. with renormalized
couplings taking some effective values, and on scales larger than quantum gravity scales ("IR"), but smaller than 
the cosmological constant ($\sim$horizon) scales ("physical"). We should use an index $r$ for renormalized couplings in 
the IR, but by an abuse of notation we will drop them. 

We will first look at the undeformed theory, with $\alpha=\beta=0$, and study $AdS_4$-type solutions in Poincare 
coordinates. In \cite{Lu:2009em}, $AdS_4$ in global coordinates was found to be a solution in the undeformed 
theory only on distance scales larger than the AdS ($\sim$horizon) size, 
i.e. for $r\sqrt{-\Lambda_W}\gg 1$, in accordance with the naive expectation above. However, below that scale
naively one still has no general relativistic invariance, and since the transformation between global and Poincare
coordinates is four dimensional, a priori it is a different calculation.
Therefore we take the ansatz
\be
ds^2=-N^2dt^2+L^2(r^2d\vec{x}^2+\frac{dr}{r^2})=L^2(-r^{2y}dt^2+r^2d\vec{x}^2+\frac{dr}{r^2})\label{adsansatz}
\ee
and keep $N$ arbitrary for the moment. Since this 
is a static solution, the $N_i$ equations of motion are satisfied. On this background, $R_{ij}=-2L^{-2}g_{ij}$
and $C_{ij}=0$. For $\alpha=\beta=0$, the $g_{ij}$ equation reduces to 
\be
\frac{\kappa^2\mu^2}{8(1-3\lambda)}
(\Lambda_w+L^{-2})Ng_{ij}\Big[\frac{3\Lambda_W-L^{-2}}{2}-L^{-2}\begin{pmatrix}y(y-1)\\0\end{pmatrix}\Big]=0
\ee
where the column corresponds to $i,j=1,2$ and $r$, respectively. Note for later use that the 0 on the position of 
$r$ would be replaced by a constraint if we replaced $g_{rr}=L^2/r^2$ with $g_{rr}=L^2/r^n$. 
The $N$ equation of motion then becomes
\be
-\frac{3\kappa^2\mu^2}{8(1-3\lambda)}[\Lambda_W^2+L^{-2}]^2=0
\ee
We see that the unique solution is $L^{-2}=-\Lambda_W$, with $N(r)$ arbitrary (in the above we have substituted 
the power law $N=r^y$, but keep $N$ general we see that it is in fact unconstrained). This is the same rather 
strange solution obtained by \cite{Lu:2009em} in global coordinates, but as we said, because of a priori 
lack of general coordinate invariance it was not obvious we should still have it. 

Turning on $\a$ and $\b$, we obtain for the $g_{ij}$ equation
\bea
&&\frac{\kappa^2\mu^2}{8(1-3\lambda)}Ng_{ij}\Big[(\Lambda_W+L^{-2})\frac{3\Lambda_W-L^{-2}}{2}+\alpha\Lambda_W L^{-2}\cr
&&-\frac{\beta}{2}\Lambda_W^2-(\Lambda_W+L^{-2}+\alpha\Lambda_W)L^{-2}\begin{pmatrix}y(y-1)\\0\end{pmatrix}\Big]=0
\label{geq}
\eea
and for the $N$ equation
\bea
&&-\frac{3\kappa^2\mu^2}{8(1-3\lambda)}[(\Lambda_W+L^{-2})^2+2\alpha L^{-2}\Lambda_W-\frac{\beta}{3}\Lambda_W^2]=0\cr
&&\Rightarrow -\frac{L^{-2}}{\Lambda_W}=1+\a\pm\sqrt{\alpha(\alpha+2)+\frac{\b}{3}}\label{neq}
\eea
Imposing that y=0 (Euclidean $AdS_3$) or 1 ($AdS_4$) are the only solutions, we get a contradiction between 
(\ref{geq}) and (\ref{neq}). Imposing instead that the coefficient of $y(y-1)$ vanishes, we get
\be
L^{-2}=-\Lambda_W(1+\alpha);\;\;\;
\beta=-3\alpha(\alpha+2)\label{adssolu}
\ee
and then it turns out that N factorizes, and N is actually arbitrary, so this is a generalization to nonzero 
$\a,\b$ of the strange solution above, so the relation between $\a$ and $\b$ is a generalization of the 
detailed balance condition. 

Note however that for $\a,\b$ arbitrary there is no Poincare 
$AdS_4$ solution at all!

We now observe that flat space, $N=1,N_i=0,g_{ij}=\delta_{ij}$, is a solution of the action if and only if
$\beta=3$, 
i.e. if there is no cosmological constant term in the action, independent of any other parameters in the action. 
In the case of the solution in (\ref{adssolu}), $L^{-2}\rightarrow
\infty$ gives a flat space limit, independent of the value of $\Lambda_W$, and we obtain $\alpha\rightarrow -1,
\beta\rightarrow 3$, i.e. both the 3d Einstein term, and the cosmological constant, vanish. However, to obtain  
the flat space solution from the start, 
only $\beta\rightarrow 3$ (no cosmological constant term) is needed, but $\a$ can be arbitrary.
Moreover, in the case $\b\rightarrow 3$, the solution of (\ref{neq}) becomes 
\be
-\frac{L^{-2}}{\Lambda_W}\simeq \frac{\b -3}{6(1+\a)}
\ee
Then, provided $y=0$ or $1$, (\ref{geq}) is also satisfied to 0th (leading) order in $L^{-2}/\Lambda_W$, 
specifically the column term vanishes
(where we should replace $g_{rr}=L^2/r^2$ with $g_{rr}=L^2/r^n$, as mentioned before) and the rest is 
negligible.

For the purpose of searching for AdS black hole solutions in the $\a,\b$ deformed theory, we will re-analyze the 
spherically symmetric ansatz considered in \cite{Lu:2009em},
\be
ds^2=-N(r)^2dt^2+\frac{dr^2}{f(r)}+r^2(d\theta^2+\sin^2\theta d\phi^2)\label{adsbh}
\ee
Following \cite{Lu:2009em}, we substitute the above ansatz with arbitrary N and f in the action and vary. We obtain the 
reduced Lagrangean
\be
{\cal L}=\frac{\kappa^2\mu^2\Lambda_W}{8(3\lambda-1)}\frac{N}{\sqrt{f}}\Big[2(1+\alpha)(1-f-rf')
+(\beta-3)\Lambda_W r^2+\frac{\lambda -1}{2\Lambda_W}f'^2-\frac{2\lambda}{\Lambda_W r^2}(f-1)f'+\frac{2\lambda-1}{
\Lambda_Wr^2}(f-1)^2\Big]
\ee
Then the $N$ equation of motion is just ${\cal L}=0$, and the $f$ equation of motion reduces to 
\be
r\d_r\ln \Big(\frac{N}{\sqrt{f}}\Big)\Big[2(1+\alpha)-\frac{\lambda-1}{\Lambda_Wr}f'+\frac{2\lambda}{\Lambda_Wr^2}
(f-1)\Big]+\frac{\lambda-1}{\Lambda_W}\Big[-f''+2\frac{f-1}{r^2}\Big]=0
\ee
We obtain that pure $AdS_4$ in global coordinates,
\be
N^2=f(r)=1-a\Lambda_Wr^2
\ee
(here $a$ corresponds to $L^{-2}/(-\Lambda_W)$ in the Poincare coordinates case)
is a solution of the $f$ equation, independent on anything else.
If $a=1+\alpha$ however, we obtain that the 
coefficient of the $N$ dependence is zero, so $N$ is actually unconstrained, as was the case at $\a=\b=0$. 

The $N$ equation gives
\bea
&&-3\Lambda_Wr^2[a^2-2a(1+\alpha)+1-\frac{\beta}{3}]=0\cr
&&\Rightarrow a=1+\a\pm \sqrt{(1+\a)^2+\frac{\beta-3}{3}}=1+\alpha\pm\sqrt{\alpha(\alpha+2)+\frac{\beta}{3}}
\eea
so just fixes $a$ in terms of $\alpha$ and $\beta$. Note that for the solution with a minus sign, the sign of $a$
is correlated with the sign of $3-\b$ (the sign of the physical cosmological constant). The solution with a plus sign
is most likely pathological.

Therefore, at arbitrary $\alpha,\beta$ we have an $AdS_4$ solution in global coordinates with 
$a$ fixed and $N^2=f$ also fixed. Thus as in the case of the deformation ${\cal L}={\cal L}_0+(1-\epsilon^2)
{\cal L}_1$ considered in \cite{Lu:2009em}, the deformation fixes the arbitrariness of $N$, while leaving the 
$AdS_4$ solution intact.

If $\beta=-3\alpha(\alpha+2)$ however, we have $a=1+\alpha$, and $N$ is arbitrary is a solution, since this is a 
generalization of the detailed balance condition.

We should not be too happy about the pure global $AdS_4$ solution at arbitrary $\a,\b$ however, since as we
saw it is not a solution in the Poincare coordinates. This is consistent, since as we discussed, if 
$\a$ and $\b$ are arbitrary and generic,
general covariance is supposed to be recovered only at large distances, which specifically means 
$|\Lambda| r^2\gg 1$ (on scales much larger than the AdS scale). Only if $\beta-3\simeq 0$ 
(i.e., $\beta-3\ll 1+\a$) we can hope to have
general covariance restored "in the physical IR" (on usual scales) as well. In that case, the cosmological constant 
of the solution is 
\be
-a\Lambda_W\simeq \frac{\b-3}{6(1+\a)}\Lambda_W\ll \Lambda_W
\ee
and $\Lambda_W$ just plays the role of one of the scales in the Newton constant, through
\be
G_N=\frac{\kappa^4\mu}{4}\sqrt{\frac{\Lambda_W(1+\a)}{3\lambda-1}}
\ee
and the existence of the (exact vs. approximate) 
solution in both global and Poincare coordinates is physically meaningful.

In general relativity, the most general rotationally invariant solution in the present of a cosmological 
constant is the AdS black hole (Birkhoff's theorem), with
\be
N^2=f=1-a\Lambda_Wr^2-\frac{M}{r}
\ee
With nonzero $M$, the equation of motion for $f$ is still satisfied by this solution.
The $N$ equation of motion then splits into terms proportional to $\Lambda_W r^2$, terms
proportional to $M/r$, which are both satisfied, and terms proportional to $M^2/(r^4\Lambda_W)$, which are left over:
\be
\frac{3}{2}(3\lambda-1)\frac{M^2}{\Lambda_Wr^4}
\ee
and thus would only vanish for $\lambda=1/3$. However, in the physical case of $\lambda=1$, or in fact at 
any $\lambda$, we conclude that we 
still have a solution (to first nontrivial order in $M$), provided we restrict ourselves to 
$r\gg (M/\Lambda_W)^{1/3}$. In the case of the real Universe, we must also choose $r\ll (a\Lambda_W)^{-1/2}$
to be on sub-horizon scales. This is the specific meaning of "in the physical IR" that we were looking for.

Note that if we had $a\sim 1$, i.e. $\beta-3\sim 1$, this condition would imply 
that the $M/r$ term is negligible with respect to $\Lambda r^2$ in the metric, which would be useless for 
gravity on anything but cosmological scales. To be useful for gravity  on usual scales, we need to fine tune 
$\beta-3\simeq 0$ to an extraordinary degree.

Until now, we have focused on static solutions, therefore we have not tested the condition of $\lambda=1$, needed 
as well in order to have four dimensional general covariance. Until now, $\lambda$ played only the role of coupling
constant, but did not affect the solutions. We have gained some undestanding of the lack of general covariance 
through the comparison of Poincare and global coordinates of AdS.

It is therefore of interest to look at solutions corresponding to objects moving at the speed of light, i.e. 
at wave solutions. In general relativity, one has the class of pp wave solutions 
\be
ds^2=2dx^+dx^-+H(x^+,x^i)(dx^+)^2+\sum_i (dx_i)^2
\ee
for which one has the 4 dimensional Ricci tensor
\be
R_{++}=-\frac{1}{2}\d_i^2 H(x^+,x^i).
\ee
For a photon, with energy-momentum tensor 
\be
T_{++}=p\delta(x^+)\delta^{d-2}(x^i)
\ee
one obtains the Aichelburg-Sexl metric \cite{Aichelburg:1970dh,Dray:1984ha}, a shockwave solution, 
\be
H(x^+,x^i)=\delta(x^+)h(x^i);\;\;\;
\Delta_{d-2}h(x^i)=-16\pi Gp \delta(x^i)
\ee
which can also be obtained by boosting a black hole of mass M to the speed 
of light, while keeping $p=e^\beta M$ fixed. It would be interesting to check under what conditions this metric
still gives a correct solution, but unfortunately the invariants $R_{ij}R^{ij}$ and $C_{ij}C^{ij}$ are nonzero 
on this solution, as we can readily check, which means that the above ansatz will contain $[\delta(x^+)]^2$ terms, 
and for this and similar reasons, it is technically difficult to check the equations of motion for such an 
ansatz.

A related solution in general relativity
is the sourceless shockwave, modelling a graviton (sourceless localized wave, travelling 
at the speed of light), satisfying $\Delta_{d-2}h(x^i)=0$, or $h(x^i)=\tilde{p}[(x_1)^2-(x_2)^2]$ in 4 dimensions.
In fact, the exact solution corresponding to the collision of two such shockwaves was found by Khan and Penrose
\cite{Khan:1971vh}.

Instead, we will look at the pp wave solutions of pure general relativity (with no source) that are also independent
of $x^+$, $H(x^+,x^i)=H(x^i)$, which are close relatives of the maximally supersymmetric pp waves of M theory and 
IIB string theory, and which have been useful in AdS/CFT as Penrose limits \cite{Berenstein:2002jq}. 
Moreover, we can think of them as being 
superpositions of the "graviton" shockwaves.

In the case of pure general relativity, 
such an ansatz will give a solution provided that $\Delta_{d-2}H(x^i)$ = $\d_i^2H=0$, i.e. 
if
\be
H(x^i)=p[(x_1)^2-(x_2)^2]\label{ppsol}
\ee
with $p$ a constant of dimension 2. While this is not a physical solution for our own Universe, 
since it does not go to flat space at infinity, it could be thought of 
as approximating a collection of gravitational shockwaves in a certain region of space. Moreover, it is a useful 
theoretical tool for understanding the limits of validity of Horava theory.

Let us then investigate the similar case in Horava theory, with 
\be
ds^2=dx^+dx^-+(dx^+)^2\phi(x^a)+\sum_{a=1,2}(dx_a)^2
\ee
and $x^\pm=y\pm ct$.
First, note that this is a solution in a flat background, so we expect it to be a solution only if $\beta=3$
(no cosmological constant).

With this ansatz, in the ADM split we have
\be
g_{yy}=1+\phi;\;\;\;
g_{ab}=\delta_{ab};\;\;\; 
N_y=-c\phi;\;\;\; N_a=0;\;\;\;
N=\frac{c}{\sqrt{1+\phi}}
\ee

We obtain the invariants
\bea
&& R_{yy}=-\frac{1}{2}\d_a^2\phi;\;\;\;
R_{ab}=-\frac{1}{2(1+\phi)}\d_a\d_b\phi+\frac{1}{4(1+\phi)^2}\d_a\phi\d_b\phi;\;\;\; R_{a\mu}=0\cr
&&R=-\frac{\d_a^2\phi}{1+\phi}+\frac{(\d_a\phi)^2}{4(1+\phi)^2}\cr
&& C_{ay}=\frac{\epsilon^{acy}}{2}\Big[\frac{\d_b \phi \d_c\d_b\phi}{4(1+\phi)}
-\frac{\d_c\phi\d_b^2\phi}{1+\phi}-\frac{1}{2}\d_c\d_b^2\phi\Big];\;\;\; C_{ab}=0\cr
&&K_{ay}=K_{ya}=\frac{\d_a\phi}{2\sqrt{1+\phi}};\;\;\;
K_{aa}=K_{yy}=0;\;\;\; K=0
\eea

Since $K=0$, it follows that $E_{ij}^{(2)}=0$. The $N_i$ equation reduces then to 
\be
\d_a^2\phi=0
\ee
as in general relativity, thus is again solved by (\ref{ppsol}).

On this solution, $R_{yy}=R_{ay}=0$, $R\sim R_{ab}\sim {\cal O}(1/X^2)$ at large $X$ (with $X$ any of the $x_a$'s),
and $K_{ij}K^{ij}\sim 1/X^2$, whereas
$C_{ay}\sim {\cal O}(p/X)$, so $Z_{ij}\sim {\cal O}(p/X;\mu w^2/X^2)$. Thus if 
\be
x\gg \kappa^2\mu\sim \frac{c}{\sqrt{\Lambda_W}};\;\;\;
p\ll \frac{w^2}{\kappa^2}\label{conds}
\ee
(where we have assumed that $\alpha$ is of order 1)
then we can ignore the higher order curvature terms in the equations of motion for $N$ and $g_{ij}$
(in the $N$ equation of motion the curvature terms are easily checked to be nonzero on the solution, so 
we need to make them small). Note that, since $\beta=3$, there is no cosmological 
constant, so in fact $\Lambda_W$ plays just the role of mass scale.

Then, since $K=0$ on the ansatz, we can have any value of 
$\lambda$, and we still obtain a solution, as in pure general relativity, 
with the only change being the renormalized value of the speed of light, 
\be
c=\frac{\kappa^2\mu}{4}\sqrt{\frac{\Lambda_W(1+\alpha)}{3\lambda -1}}
\ee
which we can then check satisfies both the $N$ equation and the $g_{ij}$ equation.

Thus the pp wave solution (\ref{ppsol}) is still a solution, independent of $\lambda$, provided we 
stay within the limit (\ref{conds}).

In conclusion, for the detailed balance action with $\a=\b=0$, or its generalization with $\b=-3\a(\a+2)$, we 
have an unusual $AdS_4$ solution (in both Poincare and global coordinates), where we can have an arbitrary 
function $N$. This case seems therefore to be pathological, but it is not surprising, since the naive expectation
is that we should only obtain general relativity on scales larger than the scale of the cosmological constant.

At general $\a,\b$ we have an exact $AdS_4$ solution in global coordinates but not in Poincare coordinates, 
reflecting the lack of general covariance in the theory.

In order to obtain an action that can approximate general relativity, we must take $\beta -3\simeq 0$ to an
extraordinary degree of precision, i.e. $|\beta-3|\ll 1+\a$. Of course, in our own Universe we have a positive 
cosmological constant (asymptotically de Sitter space), and not negative as we considered here, 
but imagining that in our Universe the cosmological constant had the same value but opposite sign, 
the principle is the same: we must have a fine tuning 
of the cosmological constant to an extraordinary degree of accuracy, like in the usual cosmological constant problem.
Note that since the action that we have used is in the IR of the theory (where all quantum corrections have been taken 
into account), this is exactly the same cosmological constant problem, just that now in the absence of any 
matter fields.

If however we have $|\beta-3|\ll 1+\a$ in the IR, 
making in effect the cosmological constant very small, independent of the 
scale $\Lambda_W$, then the Horava gravity for $S_H+\delta S_H$ will approximate general relativity at 
large distances.

{\bf Acknowledgements} 

I would like to thank Katsushi Ito for discussions.
This research has been done with partial support from MEXT's program ``Promotion of Environmental Improvement for Independence of Young Researchers'' under the Special Coordination Funds for Promoting Science and Technology, and also with partial support from MEXT KAKENHI grant nr. 20740128.

\bibliographystyle{utphys}
\bibliography{horavagrav}

\end{document}